\documentclass[12pt]{iopart}
\usepackage{epsfig,amssymb}
\def\be{\begin{equation}}
\def\ee{\end{equation}}
\def\bea{\begin{eqnarray}}
\def\eea{\end{eqnarray}}
\def\ba{\begin{array}}
\def\ea{\end{array}}
\def\nn{\nonumber}
\def\p{\partial}

\def\rr{{\quad r\to r_+\over} \!\!\!\rightarrow}

\begin{document}

\title{Hawking radiation of charged rotating AdS black holes in conformal gravity for
charged massive particles, complex scalar and Dirac particles}

\author{Gao-Ming Deng \footnote {E-mail address: denggaoming838@126.com} \\ {College of Physics and Electronic Information, \\ China West Normal University,\\
Nanchong, Sichuan 637002, China}}

\begin{abstract}
Extending researches on Hawking radiation to conformal gravity theory, we discuss Hawking radiation of different particles across
charged rotating AdS black holes in conformal gravity, including charged massive particles, complex scalar and spin-$1/2$
Dirac particles. To make the study of rotating black holes' tunneling radiation get rid of the dependence on dragging
coordinate systems, we investigate the radiation without dragging coordinate transformations. The previous geodesic
derivation existed some shortcomings. Not only did geodesics of massive and massless particles are derived by using quite
different approaches, but also the treatment for massive case was inconsistent with the variation principle of action.
Recently, Wu et al have remedied the shortcomings. In this paper, we introduce the improved treatment in conformal gravity
and derive geodesic equations of massive and massless particles in a unified and self-consistent way. Although the result
that the black holes' entropy is not one-quarter of horizon area differs from that in Einstein gravity, the tunneling
probability of charged massive particles in conformal gravity is still related to the change of Bekenstein-Hawking entropy.

\end{abstract}


%\maketitle

%\tableofcontents

\section{Introduction}

In 1970s, Hawking pointed out that black holes radiate particles thermally \cite{Hawking:1974na,Hawking:1974sw} and that the radiation
was a quantum tunneling effect owing to vacuum fluctuations near the horizons. Since then, Hawking radiation has attracted much attention \cite{Susskind:1992gd,Parikh:1999mf,Shankaranarayanan:2000gb,Zhang:2005wn,Di Criscienzo:2007fm}. Researches on it covered black holes
\cite{Hyun:1994na,Harris:2003eg}, black rings \cite{Zhao:2006zw,Miyamoto:2007ue,Chen:2007pp} and black string \cite{Peng:2008ru,Ahmed:2011fi}.
Especially the investigation of Hawking radiation from black holes involved those in de Sitter \cite{Parikh:2002qh,Medved:2002zj,Zhang:2005sf,Chen:2008ge,Farmany:2009zz,Gangopadhyay:2009hv,Zhao:2009zzj,Rahman:2012id}, anti-de
Sitter(AdS) \cite{Hemming:2000as,Podolsky:2003gm,Liu:2005hj,Wu:2006pz} space-times, static
\cite{Vagenas:2001sm,Vagenas:2002hs,Zhang:2005xt,Jiang:2005xb,Jiang:2006ea, Sarkar:2007sx,Nakayama:2008ip,Matsuno:2011ca,Kim:2011fh}
and spherical symmetric \cite{Chakraborty:2010di}, rotating charged \cite{Jiang:2005ba,Umetsu:2010kw,Wang:2009aq,Ali:2007sh},
accelerating \cite{Wu:2002ec,Zhang:2003td} and rotating \cite{Kerner:2006vu,CZ,Rehman:2010zs,Gillani:2011dj,CJZ,CYZ} black holes in
Einstein-Maxwell theory, black holes in Einstein-Maxwell-dilaton gravity \cite{Slavov:2012mv} and those in Horava-Lifshitz gravity
\cite{Chen:2009bja,Peng:2009uh,Zeng:2011zza}, as well as dynamic black holes in noncommutative gravity \cite{Mehdipour:2010ap,Nozari:2012bb}.
However, researches on Hawking radiation has not been done in conformal gravity yet. In this paper, we will extend to study Hawking
radiation of different particles across the charged rotating AdS black hole in conformal gravity \cite{Liu:2012xn}, involving charged
massive particles, the complex scalar and spin-$1/2$ Dirac particles. In fact, as a higher-derivative theory \cite{Shin:1989zh,Flanagan:2006ra},
conformal gravity deserves further explorations because of its own many intriguing properties. For example, it was initially hoped for
an alternative candidate for a renormalizable quantum gravity beyond Einstein's general relativity theory.

The derivation of particles' geodesic equations is an important aspect in the study of tunneling radiation. In previous works, there
existed some limitations in the derivation of the equations. The previous derivation of the massive particles' geodesics was inconsistent
with the variation principle of action. As for massive particles, the geodesic equation was defined by employing the relation $v_p = v_g/2$ between
the group and phase velocity. However, the variation principle demands that, in General Relativity, geodesic equations should be defined
by applying the variation principle on the Lagrangian action. On the other hand, the above treatment for massive case is not applicable
for describing the motion of massless particles. Geodesic equation of massless particles used to be derived by adopting other quite
different approach. In detail, they utilized the line element $d{s^2} = 0$ to work out massless particles' geodesic equations. In
addition, the latter derivation is of little perfection because it used inconsistent foundations-mixing together relativistic and
non relativistic descriptions. With remedying the shortcomings in mind, Wu et al \cite{Wudi:2013sc} took the Kerr black holes as an
example and made some improvements in the derivation of geodesic equations recently. Starting from the Lagrangian action and considering
the variation principle as well as three conserved integral constants, we can first obtain the geodesic equation of massive particles.
The geodesic of massless particles could be derived just by taking a proper limit of the massive case. Thus geodesic equations of massive
and massless particles can be derived in a unified and self-consistent way. Meanwhile, previous shortcomings in deriving geodesics are
overcome. Inspired by this idea, we will introduce the improved method in conformal gravity theory for the first time and derive massive
and massless particles' geodesic equations in a unified and self-consistent way.

Hawking radiation of rotating black holes was usually investigated in dragging coordinate systems. Can we study that of rotating case
without the help of dragging coordinate transformations? Actually, the principle of covariance gives us the definite and positive answer
to the question. According to the principle of covariance that all physics laws should not depend on a concrete reference system, it should
be also available to investigate that without dragging coordinate transformations, even if for the rotating case in conformal gravity theory.
Therefore, as another motivation of this paper, we will attempt to get rid of the dependence on dragging coordinate systems so that rotating
black holes' tunneling radiation could be discussed in a general non-dragging coordinate system and extend that to conformal gravity. In this
paper, the tunneling probability of the charged rotating AdS black holes \cite{Liu:2012xn} in conformal gravity will be calculated in a
non-dragging coordinate system.

Many methods can be adopted to study Hawking radiation as tunneling, such as Parikh-Wilczek's semi-classical tunneling method \cite{Parikh:1999mf} 
(originated from the early developments \cite{Kraus:1994by,Kraus:1994fj,KeskiVakkuri:1996xp}), the Hamilton-Jacobi tunneling method \cite{Angheben:2005rm} 
which is developed from the complex-path analysis proposed by Padmanabhan et al \cite{Srinivasan:1998ty,Shankaranarayanan:2000qv}. For these three 
kinds of particles, we will study their Hawking radiation by using the Parikh-Wilczek's semi-classical tunneling method and the methods resorting
to Klein-Gordon and Dirac equations. It is noteworthy that, in order to work out tunneling probabilities of the complex scalar and spin-$1/2$
Dirac particles, we just need to deal with the radial parts of Klein-Gordon and Dirac equations while disregarding the angular parts after
variables separation. As we will see later, this improving operation provides a convenience for simplifying derivation process in comparison
with previous works.

Our paper is motivated by extending researches on Hawking tunneling radiation to conformal gravity theory with some improvements in deriving
particles' geodesics and avoiding dragging coordinate transformations in order to investigate the tunneling radiation of rotating black holes
in general non-dragging coordinate systems. The topic of this paper is focused on discussing the tunneling radiation of recently constructed
charged rotating AdS black hole \cite{Liu:2012xn} in conformal gravity.

The remainder of this paper is organized as follows. We will first review the charged rotating AdS black hole solution \cite{Liu:2012xn} in the
context of conformal gravity and its relevant thermodynamical aspect in section \ref{NFCRAB}. In sections \ref{TPOCMPB}, \ref{TPOCSP} and
\ref{TPOCDP}, the tunneling radiation of charged massive particles, complex scalar and spin-$1/2$ Dirac particles will be investigated by using
the Parikh-Wilczek's semi-classical tunneling method, the methods resorting to Klein-Gordon and Dirac
equations. With remedying previous shortcomings in deriving geodesic equations, the geodesic equations of massive and massless tunneling particles
will be derived from the Lagrangian action in a unified and self-consistent way in subsection \ref{GEOCMP}. Section \ref{CADA} is devoted to
conclusions and discussion.

%%%%%%%%%%%%%%%%%%%%%%%%%%%%%%%%%%%%%%%%%%%%%%%%%%%%%%%%%%%%%%%%%%%%%%
\section{Charged rotating AdS black hole solution in conformal gravity and its thermodynamics}
%%%%%%%%%%%%%%%%%%%%%%%%%%%%%%%%%%%%%%%%%%%%%%%%%%%%%%%%%%%%%%%%%%%%%%
\label{NFCRAB}

As an intriguing higher-derivative theory, conformal gravity has attracted more and more interest. Recently, L\"{u} et al obtained a new charged
rotating AdS black hole solution \cite{Liu:2012xn} in conformal gravity. The black hole solution takes the analogous form as charged Kerr-Newman-AdS
solution in Boyer-Lindquist-type coordinates in Einstein gravity \cite{Carter:1968ks,Plebanski:1976gy} except for the difference in the value of
${\Delta_r}$. We will first review the new charged rotating AdS black hole \cite{Liu:2012xn} in conformal gravity and the thermodynamic properties.
Introducing coordinate transformations $t \rightarrow \tilde{t}$, $\phi \rightarrow \tilde{\phi} -ag^2\tilde{t}$, the charged rotating AdS black
hole solution in four dimensions and the vector potential can be written as
\bea
d\tilde{s}^2 &=& -\frac{\Delta_r}{\Sigma\Xi^2}\left(\Delta_\theta d\tilde{t} -a\sin^2\theta d\tilde{\phi}\right)^2 +\Sigma\Big(\frac{dr^2}{\Delta_r}
+\frac{d\theta^2}{\Delta_\theta}\Big) \nn \\
&& +\frac{\Delta_\theta \sin^2\theta}{\Sigma\Xi^2}\Big[a(1 +g^2r^2)d\tilde{t} -(r^2 +a^2)d\tilde{\phi}\Big]^2,
\label{formula01}
\eea
\be
\tilde{\mathcal{A}} = \frac{{qr}}{{\Sigma \Xi }}\left( {{\Delta _\theta }d\tilde t - a{{\sin }^2}\theta d\tilde \phi } \right)\, ,\
\label{formula02}
\ee
where
\bea
&& \Sigma = r^2 +a^2\cos^2\theta \, , \quad \Delta_{\theta} = 1 -g^2a^2\cos^2\theta, \nn \qquad~\\
&& \Xi = 1 -g^2a^2,\quad \Delta_r = (r^2 +a^2)(1 +g^2r^2) -2m r +\frac{{{q^2}{r^3}}}{{6m}}, \nn \qquad~
\eea
in which $a$ and $q$ are rotating parameter and charge parameter, $m$ is an integration constant related to the mass, also ${g^2}$ has connection
with the cosmological constant $\Lambda = - 3{g^2}$.

Accompanying with the charged rotating AdS black hole solution, corresponding thermodynamical quantities were also given in Ref.\cite{Liu:2012xn}.
First, the entropy and temperature are given by
\be
\hspace{-1.5cm} T = \frac{{3(1 + 3{g^2}r_ + ^2)r_ + ^2 - 3{a^2}(1 - {g^2}r_ + ^2) + \frac{{{q^2}}}{m}r_ + ^3}}{{12\pi {r_ + }(r_ + ^2 + {a^2})}}\, , \quad
S = \frac{{2\alpha \pi }}{\Xi }\big(1 + {g^2}r_ + ^2 + \frac{{{q^2}{r_{\rm{ + }}}}}{{6m}}\big),
\label{formula03}
\ee
in which $\alpha $ is a coupling constant. The electric charge and corresponding electric potential are
\be
Q = - \frac{{\alpha q}}{{3\Xi }}\, , \quad \Phi = \frac{{q{r_ + }}}{{r_ + ^2 + {a^2}}}\, .
\label{formula04}
\ee
The angular velocity at the horizon and the angular momentum of the black hole are computed as
\be
\Omega = \frac{{a(1 + {g^2}r_ + ^2)}}{{r_ + ^2 + {a^2}}}\, , \quad J = \frac{{2\alpha a{g^2}}}{{{\Xi ^2}}}\left( {m + \frac{{{q^2}}}{{12m{g^2}}}} \right)\, .
\label{formula05}
\ee
Moreover, the mass and thermodynamical potential $\Theta$ conjugated to the cosmological constant $\Lambda $ are given by
\be
E = \frac{{2\alpha {g^2}}}{{{\Xi ^2}}}\left( m + \frac{a^2q^2}{12m} \right)\, , \quad
\Theta = - \frac{\alpha(r_+^2 +{a^2})}{6{\Xi^2}{r_+}}\left(1 + {g^2}r_+^2 + \frac{{q^2}{r_+}}{6m}\right)\, .
\label{formula06}
\ee

It is verified that the differential form of the first law of black hole thermodynamics holds
\be
dE = TdS + \Omega dJ + \Phi dQ + \Theta d\Lambda \, ,
\label{formula07}
\ee
while the Smarr formula takes the following unusual form
\be
E = 2\Theta \Lambda \, .
\label{formula08}
\ee
It can be verified that the formula (\ref{formula08}) is indeed valid despite it takes unusual form. The fact that the usual integral Smarr formula
can not hold true anymore in conformal gravity theory might hint an important difference \cite{Lu:2012xu} of conformal gravity from that of the
General Relativity.

With these thermodynamic properties in hand, we will proceed to study Hawking radiation of the charged rotating AdS black holes in conformal gravity.
The previous research on Hawking radiation from rotating black holes usually needed the help of dragging coordinate transformations. But according to
the principle of covariance, it's also available to investigate that in non-dragging coordinate systems, even if in conformal gravity theory. We will
next attempt to get rid of the dependence on dragging coordinate transformations to discuss the tunneling radiation from the rotating black holes
\cite{Liu:2012xn} in a general non-dragging coordinate system. In order to calculate the tunneling probabilities of these three kinds of particles, we
will use different methods such as the Parikh-Wilczek's semi-classical tunneling method, the methods resorting to Klein-Gordon and Dirac equations. 
First of all, we begin with studying the tunneling radiation of charged massive particles.

%%%%%%%%%%%%%%%%%%%%%%%%%%%%%%%%%%%%%%%%%%%%%%%%%%%%%%%%%%%%%%%%%%%%%%%%%%%%%%%%%%%%%
\section{Tunneling radiation of charged massive particles}
%%%%%%%%%%%%%%%%%%%%%%%%%%%%%%%%%%%%%%%%%%%%%%%%%%%%%%%%%%%%%%%%%%%%%%%%%%%%%%%%%%%%%
\label{TPOCMPB}

In this section, Hawking radiation via tunneling of charged massive particles is discussed by using the Parikh-Wilczek's semi-classical tunneling method.
We first step from deriving the geodesic equations of particles.

%%%%%%%%%%%%%%%%%%%%%%%%%%%%%%%%%%%%%%%%%%%%%%%%%%%%%%%%%%%%%%%%%%%%%%%%%
\subsection{Geodesic equations of charged massive particles}
%%%%%%%%%%%%%%%%%%%%%%%%%%%%%%%%%%%%%%%%%%%%%%%%%%%%%%%%%%%%%%%%%%%%%%%%%
\label{GEOCMP}

Previous related works derived geodesic equations of massive and massless particles by adopting quite different methods: as for massless particles they
utilized the line element $d{s^2} = 0$ to define the geodesics, but for massive particles they derived the geodesic equation resorting to the relation
$v_p = v_g/2$ between the group and the phase velocity. Not only did the former treatment mix together relativistic and non relativistic descriptions
in the process of deriving the massless geodesics, but also the latter approach was inconsistent with the variation principle and was quite different from
the derivation of the massless case. To remedy the shortcomings, both the null and timelike geodesic equations can be derived from the Lagrangian action
with the help of the variation principle actually. As for massive and massless particles' geodesics in black holes, there exist three conserved integral
constants including energy $\mathcal{E}$ and angular momentum $L$ which correspond to Killing vector $\p_t,\p_\phi$ respectively as well as the Hamiltonian
$\mathcal{H}$ which can be restricted to be a constant $\mathcal{H} = -m_0K/2$, ($K = 0, 1$). It should be pointed out that the constant $K$ is permissible
to take two different values $0, 1$ corresponding to the 4-velocity normalization condition of null and timelike geodesic, respectively. These integral
constants are enough to determine the motion equations of particles. We can first use these integral constants of motion to derive geodesic equations of
massive tunneling particles. The massless particles' geodesic equation could be obtained just by taking a proper limit of the massive case. So geodesic
equations of massive and massless particles can be derived in a unified and self-consistent way. In this subsection, with the improvements, we shall take
advantage of unified and self-consistent method to derive geodesic equations of test particles in conformal gravity for the first time.

First, by performing following coordinate transformation
\bea
d\tilde{t} =& dt -\frac{\sqrt{r^2 +a^2}\sqrt{(r^2 +a^2\big)\big(1 +g^2r^2)-\Delta_r}}{\Delta_r\sqrt{1 +g^2r^2}}dr \, , \qquad \\
\label{formula09}
d\tilde{\phi} =& d\phi -\frac{a\sqrt{1 +g^2r^2}\sqrt{(r^2 +a^2)(1 +g^2r^2)-\Delta_r}}{\Delta_r\sqrt{r^2 +a^2}}dr \, ,
\label{formula10}
\eea
the line element (\ref{formula01}) and the vector potential (\ref{formula02}) are changed to the following form,
\bea
\hspace{-2.6cm} ds^2 &&= -\frac{(1 +g^2r^2)\Delta_\theta}{\Xi}dt^2 +\frac{\Sigma}{\Delta_\theta}d\theta^2
 +\frac{(r^2 +a^2)\sin^2\theta}{\Xi}d\phi^2 \nn \\
\hspace{-2.6cm} && +\bigg[\frac{\sqrt{(r^2 +a^2)(1 +g^2r^2) -\Delta_r}}{\sqrt{\Sigma}\Xi}(\Delta_\theta dt
 -a\sin^2\theta d\phi) +\sqrt{\frac{\Sigma}{(r^2 +a^2)(1 +g^2r^2)}}dr\bigg]^2,
\label{formula11}
\eea
\bea
\mathcal{A} &=& \frac{qr}{\Sigma \Xi}\bigg({\Delta_\theta}dt -a{{\sin }^2}\theta d\phi \bigg) \, ,
\label{formula12}
\eea
in which $\Sigma,{\Delta _ r},\Xi,{\Delta _\theta }$ take the same values as that in the metric (\ref{formula01}). It is obvious that in the new
form (\ref{formula11}), the metric is well behaved at the horizon.

Considering a charged massive particle, the corresponding Lagrangian $\mathcal{L} = m_0 g_{\mu\nu} \dot{x}^{\mu}\dot{x}^{\nu}/2 +\mathcal{Q} \mathcal{A}_{\mu}\dot{x}^{\mu}$ is
\bea
\mathcal{L} &=& \frac{m_0}{2} \Bigg\{-\frac{(1 +g^2r^2)\Delta_\theta}{\Xi}\dot{t}^2
 +\frac{\Sigma}{\Delta_\theta}\dot{\theta}^2 +\frac{(r^2 +a^2)\sin^2\theta}{\Xi}\dot{\phi}^2 \nn \\
&& +\bigg[\frac{\sqrt{(r^2 +a^2)(1 +g^2r^2)
 -\Delta_r}}{\sqrt{\Sigma}\Xi}(\Delta_\theta \dot{t} -a\sin^2\theta \dot{\phi}) \nn \\
&& +\dot{r}\sqrt{\frac{\Sigma}{(r^2 +a^2)(1 +g^2r^2)}}\bigg]^2 \Bigg\}
 +\frac{\mathcal{Q}q r}{\Sigma \Xi}\bigg({\Delta_\theta}\dot{t} -a{{\sin }^2}\theta \dot{\phi}\bigg) \, .
\qquad
\label{formula13}
\eea
in which $m_0,{\mathcal{Q}}$ are the rest mass and electric charge of particle respectively. Subsequently, the generalized
momentum $P_\alpha = \p\mathcal{L}/\p\dot{x}^{\alpha}$  can be deduced,
\bea
P_t &=& \frac{m_0\Delta_\theta}{\Xi} \Bigg\{-(1 +g^2r^2)\dot{t} +\frac{\mathcal{Q}q r}{m\Sigma}
+\dot{r}\sqrt{1 -\frac{\Delta_r}{(r^2 +a^2)(1 +g^2r^2)}} \nn \\
&& +\frac{(r^2 +a^2)(1 +g^2r^2) -\Delta_r}{\Sigma\Xi}(\Delta_\theta\dot{t}
-a\sin^2\theta\dot{\phi}) \Bigg\}, \qquad \\
 \label{formula14}
P_r &=& m_0\bigg[\frac{\Sigma}{(r^2 +a^2)(1 +g^2r^2)}\dot{r} \nn \\
&& +\sqrt{1 -\frac{\Delta_r}{(r^2 +a^2)(1 +g^2r^2)}}
 \frac{\Delta_\theta\dot{t} -a\sin^2\theta\dot{\phi}}{\Xi} \bigg], \qquad \label{formula15} \\
P_\theta &=& \frac{m_0\Sigma}{\Delta_\theta}\dot{\theta} \, , \label{formula16} \\
P_\phi &=& \frac{m_0\sin^2\theta}{\Xi} \Bigg\{(r^2 +a^2)\dot{\phi} -\frac{\mathcal{Q}q r a}{m\Sigma} \nn \\
&& -a \bigg[\frac{(r^2 +a^2)(1 +g^2r^2) -\Delta_r}{\Sigma\Xi}(\Delta_\theta\dot{t} -a\sin^2\theta\dot{\phi}) \nn \\
&& +\dot{r}\sqrt{1 -\frac{\Delta_r}{(r^2 +a^2)(1 +g^2r^2)}}\bigg]\Bigg\}.
\label{formula17}
\eea
With the help of the Legendre transformation, we can get the Hamiltonian as follows
\bea
\hspace{-2.5cm}\mathcal{H} &=& \dot{t}P_t +\dot{r}P_r +\dot{\theta}P_\theta +\dot{\phi}P_\phi -\mathcal{L} \nn \\
\hspace{-2.5cm} &=& \frac{m_0}{2} \Bigg\{-\frac{(1 +g^2r^2)\Delta_\theta}{\Xi}\dot{t}^2
 +\frac{\Sigma}{\Delta_\theta}\dot{\theta}^2 +\frac{(r^2 +a^2)\sin^2\theta}{\Xi}\dot{\phi}^2 \nn \\
\hspace{-2.5cm} && +\bigg[\frac{\sqrt{(r^2 +a^2)(1 +g^2r^2)
 -\Delta_r}}{\sqrt{\Sigma}\Xi}(\Delta_\theta \dot{t} -a\sin^2\theta \dot{\phi})
+\dot{r}\sqrt{\frac{\Sigma}{(r^2 +a^2)(1 +g^2r^2)}}\bigg]^2 \Bigg\},
\label{formula18}
\eea
in which the variables $t$ and $\phi$ act as cyclic coordinates that correspond to the conserved generalized momenta $P_t$ and $P_\phi$.
As analysis at the beginning of this section, let $P_t,P_\phi$ be respectively the integral constants $\mathcal{E}$ and $L$ and consider
the 4-velocity normalization condition $\mathcal{H} = -m_0 K/2$, ($K = 0, 1$), that is,
\be
P_t = \mathcal{E} \, , \quad P_\phi = L \, , \quad \mathcal{H} = -m_0 K/2 \, . \label{formula19}
\ee

Using the above three equations and (\ref{formula16}), we get the geodesic equations of charged massive particles. The radial part is
\bea
\hspace{-2.5cm} \bar{r} &=& \frac{dr}{dt} = \frac{\dot{r}}{\dot{t}} \nn \\
\hspace{-2.5cm} &=& \Delta_r\Bigg[(r^2 +a^2)\sqrt{1 -\frac{\Delta_r}{(r^2 +a^2)(1 +g^2r^2)}}
 \pm \frac{-(r^2 +a^2)Y +a\Delta_r\big(L +\frac{a\mathcal{E}\sin^2\theta}{\Delta_\theta}\big)}{\sqrt{W}}\Bigg]^{-1},
\label{formula20}
\eea
and the angular part is
\bea
\hspace{-1.8cm} \bar{\phi} &=& \frac{d\phi}{dt} = \frac{\dot{\phi}}{\dot{t}} \nn \\
\hspace{-1.8cm} &=& \frac{a\bigg(1 +g^2r^2\bigg)\Bigg\{-Y +\frac{\Delta_r\Delta_\theta\Big(L +\frac{a\mathcal{E}\sin^2\theta}{\Delta_\theta}\Big)}
 {a\Big(1 +g^2r^2\Big)\sin^2\theta} \pm \sqrt{\Big[1 -\frac{\Delta_r}{\big(r^2 +a^2\big)\big(1 +g^2r^2\big)}\Big]W}\Bigg\}}{-\bigg(r^2 +a^2\bigg)Y +a\Delta_r\Big(L +\frac{a\mathcal{E}\sin^2\theta}{\Delta_\theta}\Big)
 \pm \sqrt{\Big[r^2 +a^2 -\frac{\Delta_r}{1 +g^2r^2}\Big]\left(r^2 +a^2\right)W}}\, , \quad
\label{formula21}
\eea
where
\bea
 Y &=& (r^2 +a^2)\mathcal{E} +a L(1 +g^2r^2) -\mathcal{Q}q r ,\qquad \nn \\
 W &=& Y^2 -\Delta_r\bigg[\frac{(L\Delta_\theta +a\mathcal{E}\sin^2\theta)^2}{\Delta_\theta\sin^2\theta}
 -m_0^2K\Sigma -\Delta_\theta P_\theta^2\bigg]\, . \nn
\eea
and $``\pm"$ correspond to the geodesics of ingoing and outgoing particles. The geodesic equation of massless particles could be obtained
by taking a proper limit of the massive case. Therefore, the geodesic equations of massive and massless particles in conformal gravity are
derived in a unified and self-consistent way.

In addition, we will investigate the asymptotic behaviors of the outgoing particles near the event horizon. When $r$ goes to the horizon radius
$r_+$, the radial equation (\ref{formula20}) and angular equation (\ref{formula21}) behave like
\bea
\bar{r} &=& \frac{dr}{dt} \rr \frac{\Delta^{\prime}(r_+)(r -r_+)}{2(r_+^2 +a^2)} = \kappa(r -r_+), \qquad~ \\
\label{formula22}
\bar{\phi} &=& \frac{d\phi}{dt} \rr \frac{a(1 +g^2r_+^2)}{r_+^2 +a^2}, \label{formula23}
\eea
where $\Delta^{\prime}(r_+)$ denotes the first derivative of $\Delta_r$ at the horizon radius $r_+$ and $\kappa$ is the surface gravity on the
horizon.

%%%%%%%%%%%%%%%%%%%%%%%%%%%%%%%%%%%%%%%%%%%%%%%%%%%%%%%%%%%%%%%%%%%%%%%%%
\subsection{Tunneling probability of charged massive particles}
%%%%%%%%%%%%%%%%%%%%%%%%%%%%%%%%%%%%%%%%%%%%%%%%%%%%%%%%%%%%%%%%%%%%%%%%%
\label{TPOCMPBUT}

According to the process of Hawking radiation, a pair of particles is produced near the horizon due to vacuum fluctuations, the positive energy
particle tunnels out to the infinity while the negative energy ``partner" is absorbed into the black hole. The energy of the black hole decreases
and the horizon shrinks. Here the self-gravitation is taken into account. We consider the conservation of the energy and angular momentum and introduce
a new constant $\hat{q}$ satisfied $q = m \hat{q}$ in (\ref{formula11}). When a particle with the energy $\mathcal{W}$ and the charge $\mathcal{Q}$
tunnels out the black hole, the mass, angular momentum and electric charge of the black hole would change to $\big( {M - \mathcal{W}} \big)$,
$\frac{a\big(12 g^2 + {\hat{q}}^2\big)\big(M - \mathcal{W} \big)}{{g^2}\big(12 + a^2{\hat{q}}^2\big)}$ and $\big( {Q - \mathcal{Q}} \big)$,
respectively. Then the line element and the vector potential become
\bea
\hspace{-2.5cm} d\bar{s}^2 &&= -\frac{(1 +g^2r^2)\Delta_\theta}{\Xi}dt^2 +\frac{\Sigma}{\Delta_\theta}d\theta^2
 +\frac{(r^2 +a^2)\sin^2\theta}{\Xi}d\phi^2 \nn \\
\hspace{-2.5cm} && \hspace{-0.3cm} +\bigg[\frac{\sqrt{\big(r^2 +a^2\big)\big(1 +g^2r^2\big) -\bar{\Delta}_r}}{\sqrt{\Sigma}\Xi}(\Delta_\theta dt
 -a\sin^2\theta d\phi) +\sqrt{\frac{\Sigma}{(r^2 +a^2)(1 +g^2r^2)}}dr\bigg]^2,
\label{formula24}
\eea
\be
\bar{\mathcal{A}} = -\frac{3\Xi(Q -\mathcal{Q})}{\alpha \Sigma\Xi}\left({\Delta_\theta}dt -a{{\sin }^2}\theta d\phi\right), \qquad~~
\label{formula25}
\ee
where ${\bar{\Delta}_r} = ({r^2} + {a^2})(1 + {g^2}{r^2}) - \frac{{{\Xi ^2}r\big(12 - {{\hat{q}}^2}{r^2}\big)
\big(M - \mathcal{W} \big)}}{{\alpha {g^2}\big(12 + {a^2}{{\hat{q}}^2}\big)}}$.

According to the WKB approximation, the tunneling probability is derived by the action of the particle, which is
\be
\Gamma\sim e^{-2ImS} \, . \label{formula26}
\ee
To proceed with an explicit calculation, we adopt the Hamilton equations
\be
\bar{r} = \frac{d\mathcal{H}}{dP_r}\Big|_{(r; \phi,P_\phi,\mathcal{A}_t,P_{\mathcal{A}_t})}
= \frac{d(M -\mathcal{W})}{dP_r} \, ,
\label{formula27}
\ee
\be
\bar{\phi} = \frac{d\mathcal{H}}{dP_\phi}\Big|_{(\phi; r,P_r,\mathcal{A}_t,P_{\mathcal{A}_t})}
           = \frac{a\Omega \big(12 g^2 + {\hat{q}}^2\big)}{g^2 \big(12 + a^2{\hat{q}}^2\big)}\frac{d(M -\mathcal{W})}{dP_\phi} \, ,
\label{formula28}
\ee
\be
\bar{\mathcal{A}_t} = \frac{d\mathcal{H}}{dP_{\mathcal{A}_t}}\Big|_{(\mathcal{A}_t; r,P_r,\phi,P_\phi)}
= \Phi\frac{d(Q -\mathcal{Q})}{dP_{\mathcal{A}_t}} \, ,
\label{formula29}
\ee
where we have utilized equations $d\mathcal{H}_{(\phi;r,P_r,\mathcal{A}_t,P_{\mathcal{A}_t})} = \Omega dJ
= \frac{a\Omega \big(12 g^2 + {\hat{q}}^2\big)}{g^2 \big(12 + a^2{\hat{q}}^2\big)} d\big(M -\mathcal{W}\big)$
and $P_\phi = J =\frac{a\big(12 g^2 + {\hat{q}}^2\big)\big(M -\mathcal{W}\big)}{g^2 \big(12 + a^2{\hat{q}}^2\big)}$.
Substituting the action (\ref{formula18}) into Eqs. (\ref{formula27})-(\ref{formula29}) yields
\be
\hspace{-1.8cm} S = \int_{t_i}^{t_f}\big(P_r\bar{r} -P_{\phi}\bar{\phi} -P_{\mathcal{A}_t}\bar{\mathcal{A}_t}\big)dt
 = \int\limits_{r_i}^{r_f}\Bigg[\hspace{-0.1cm}\int\limits_{(0, ~0, ~0)}^{(P_r, P_{\phi}, P_{\mathcal{A}_t})}
\hspace{-0.3cm} \Big(\bar{r}~dP_r^{\prime} -\bar{\phi}~dP_{\phi}^{\prime}
-\bar{\mathcal{A}_t}~dP_{\mathcal{A}_t}^{\prime}\Big)\Bigg]\frac{dr}{\bar{r}}, \quad~
\label{formula30}
\ee
where, $r_i$ and $r_f$ correspond to the horizon radius before and after the shrinkage, $P_r$ and $P_\phi$ are the canonical momenta
conjugated to $r$ and $\phi$, respectively. Eventually, we have
\be
\hspace{-2.5cm}\textrm{Im}~ S = \textrm{Im} \int_{r_i}^{r_f}\int_{(M, ~Q)}^{(M -\mathcal{W}, ~Q -\mathcal{Q})}
\Big\{\left[1 -\frac{a\Omega^{\prime} (12 g^2 + {\hat{q}}^2)}{g^2 (12 + a^2{\hat{q}}^2)}\right]
d(M -\mathcal{W}^{\prime}) -\Phi^{\prime}d(Q -{\mathcal{Q}}^{\prime})\Big\}\frac{dr}{\bar{r}}. \nn \\
\label{formula31}
\ee

To simplify the calculation of the above integral, we make use of the asymptotic behavior of the outgoing particles near the
event horizon
\be
\bar{r} \approx \kappa^{\prime}\,(r -{r}^{\prime}_+) \, , \qquad
\kappa^{\prime}= \frac{{\bar{\Delta}}^{\prime}({r}^{\prime}_+)}{2({{r}^{\prime}_+}^2 +a^2)} \, ,
\label{formula32}
\ee
where $\kappa^{\prime}$ is the surface gravity on the horizon after the particle's emission and ${\bar{\Delta}}^{\prime}({r}^{\prime}_+)$ is
the first derivative of ${\bar \Delta _r}$ at horizon radius ${r}^{\prime}_+$. What's more, the angular velocity and the electrostatic potential
are given by
\be
{\Omega^{\prime}_ +} = \frac{a}{{{r^{\prime}_+}^2 + {a^2}}}\, , \qquad
{\Phi^{\prime}_+} = \frac{(Q - \mathcal{Q}^\prime) r^{\prime}_+}{{r^{\prime}_+}^2  + {a^2}} \, ,
\label{formula33}
\ee
It can be verified that these thermodynamical quantities comply with the differential form of the first law of black hole thermodynamics
when $\Lambda = -3g^2$ is treated as a constant, namely,
\be
\hspace{-0.8cm} d\big(M -\mathcal{W}^{\prime}\big) = \frac{\kappa^{\prime}}{2\pi}d S^{\prime}  +\frac{a\Omega^{\prime} (12 g^2 + {\hat{q}}^2)}{g^2 (12 + a^2{\hat{q}}^2)}d(M -\mathcal{W}^{\prime}) +\Phi^{\prime}_+ d(Q -{\mathcal{Q}}^{\prime}).
\label{formula34}
\ee
Substituting Eqs. (\ref{formula32}) and (\ref{formula34}) into (\ref{formula31}) and using the equation (\ref{formula33}), the
imaginary part of the action is obtained as follows
\bea
\hspace{-1.8cm}\textrm{Im}S&\approx& \textrm{Im} \int_{r_i}^{r_f}\int_{(M, ~Q)}^{(M -\mathcal{W}, ~Q -\mathcal{Q})}
\frac{\bigg[1 -\frac{a\Omega^{\prime}_+ (12 g^2 + {\hat{q}}^2)}{g^2 (12 + a^2{\hat{q}}^2)}\bigg]
 d(M -\mathcal{W}^{\prime}) -\Phi^{\prime}_+ d(Q -{\mathcal{Q}}^{\prime})}{\kappa^{\prime}(r -r_+^{\prime})} dr \nn \\
\hspace{-1.8cm}&=& -\frac{1}{2} \int_{S_{BH}(M, ~Q)}^{S_{BH}(M -\mathcal{W}, ~Q -\mathcal{Q})} dS^{\prime} = -\frac{1}{2}\Delta S_{BH}\, .
\label{formula35}
\eea
Consequently, the tunneling probability of the particles is
\be
\Gamma\sim e^{-2\textrm{Im} S} = e^{\triangle S_{BH}} \, . \label{formula36}
\ee
It is noteworthy that, the entropy (\ref{formula03}) of the black holes in conformal gravity is not one-quarter of the horizon area, but
nevertheless the tunneling probability is still related to the change of Bekenstein-Hawking entropy. In this paper, the investigation of
the rotating black hole's tunneling radiation was carried out in a non-dragging coordinate system.

%%%%%%%%%%%%%%%%%%%%%%%%%%%%%%%%%%%%%%%%%%%%%%%%%%%%%%%%%%%%%%%%%%%%%%%%%%%%%%%%%%%%%
\section{Tunneling radiation of complex scalar particles}
%%%%%%%%%%%%%%%%%%%%%%%%%%%%%%%%%%%%%%%%%%%%%%%%%%%%%%%%%%%%%%%%%%%%%%%%%%%%%%%%%%%%%
\label{TPOCSP}

In this section, we investigate the tunneling radiation of charged complex scalar particles by using the method which resorted to the 
Klein-Gordon equation.

In curved spacetime, the Klein-Gordon equation is written as
\be
\frac{1}{\sqrt{-g}}\left[ \left( \frac{\p}{\p{x^\mu}}
+ ie{\tilde{\mathcal{A}}_\mu} \right)\sqrt{-g} {g^{\mu \nu }} \left( \frac{\p}{\p{x^\nu}}
+ ie{\tilde{\mathcal{A}}_\nu}\right)\Phi \right] -{\mu_0}^2\Phi = 0,
\label{formula37}
\ee
in which $\mu_0 $ is the mass of the scalar particle and $e$ is the charge. Applying the inverse metric in (\ref{formula01}), the Klein-Gordon
equation is
\bea
\hspace{-0.8cm} && -\frac{1}{\Sigma {\Delta_r}} \left[ (r^2 +a^2)\frac{\p}{\p t} +a(1 + g^2r^2)\frac{\p}{\p \phi}
+ e q r \right]^2\Phi +\frac{1}{\Sigma }\frac{\p}{\p r}({\Delta_r}\frac{\p\Phi}{\p r}) \nn \\
\hspace{-0.8cm} &&  + \frac{1}{\Sigma {\sin\theta}}\frac{\p}{\p\theta}({\Delta_\theta}{\sin\theta} \frac{\p\Phi}{\p\theta})
 +\frac{1}{\Sigma \Delta_\theta}(a{\sin\theta} \frac{\p}{\p t} + \frac{\Delta_\theta}{\sin\theta}\frac{\p}{\p\phi})^2\Phi  - {\mu_0}^2\Phi = 0.
\label{formula38}
\eea
To solve the above equation, we carry out separation of variables
\be
\Phi \big(t,r,\theta,\phi \big) = {e^{-i\omega_1 t}}{e^{im_1\phi}}X\left(\theta\right)R\left(r\right).
\label{formula39}
\ee
where $\omega_1$ denotes the energy of the radiation particle, $m_1$ is the projection of the angular momentum of the radiation particle on the
rotation axis. Inserting the separation into the Klein-Gordon equation yields the radial equation
\be
\hspace{-2.5cm} \Big\{-\frac{1}{\Delta_r}\left[\omega_1 \left(r^2 +a^2\right) -am_1\left(1 +g^2r^2\right) +e q r\right]^2
 +\xi  +{\mu_0}^2{r^2}\Big\} R(r) = \frac{d}{dr}\Big[{\Delta_r}\frac{d}{dr}R(r)\Big],
\label{formula40}
\ee
and the angular part
\be
\hspace{-2.0cm}\left[\frac{1}{\Delta_\theta}{{\big( a\omega_1 {\sin\theta}  -\frac{m{\Delta_\theta}}{\sin\theta} \big)}^2} + a^2{\mu_0}^2{{\cos^2}\theta} -\xi \right]X(\theta)
= \frac{1}{\sin\theta}\frac{d}{d\theta}\Big[{\Delta_\theta}{\sin\theta} \frac{d X\left(\theta\right)}{d\theta}\Big],
\label{formula41}
\ee
where $\xi$ is constant. To proceed with calculating the tunneling probability of particles, we assume
\be
R(r)= Ce^{\frac{i}{\hbar}{W\left( r \right)}}\, . \quad
\label{formula42}
\ee
Following with recovering $\hbar$ and replacing $m_1,\omega_1$ with $\frac{\mathcal{J}}{\hbar},\frac{\mathcal{E}}{\hbar}$, equation
(\ref{formula40}) turns to
\bea
\hspace{-0.5cm}
&& -\frac{1}{\Delta_r}\{\left[ \mathcal{E}\left( r^2 + a^2 \right) -
a\mathcal{J}\left( 1 +g^2r^2 \right) + e q r \right]^2 +\xi +{\mu_0}^2 r^2\} \nn \\
\hspace{-0.5cm}
&& = i[\Delta_r^\prime W^\prime\left( r \right) +\frac{i}{\hbar} \Delta_r W^\prime\left( r \right)^2 +\Delta_r W''\left( r \right)] \, ,
\label{formula43}
\eea
in which $\Delta_r^{\prime},W''\left( r \right)$ denote the first derivative of $\Delta_r$ and the second derivative
of $W(r)$. After discarding high order terms and taking the following near-horizon approximation into consideration 
\be
\hspace{-1.5cm} \Delta_r = (r -r_+)\Delta^{\prime}(r_+) +\cdots \, , \quad
\Omega_+ = \frac{{a(1 + {g^2}r_ + ^2)}}{{r_ + ^2 + {a^2}}}\, , \quad
\kappa= \frac{{\Delta}^{\prime}(r_+)}{2(r_+^2 +a^2)}\, , \quad
\label{formula44}
\ee
we solve the equation for
$W^\prime\left( r \right)$ and get
\be
W_\pm^{\prime}(r) = \pm \frac{1}{2\kappa \left( r -r_+ \right)}\left(\mathcal{E} -{\Omega_+}\mathcal{J}
+ \frac{e q {r_+}}{r_+^2 + a^2} \right).
\label{formula45}
\ee
Integrating the above equation yields
\be
{W_ \pm }(r) =  \pm \frac{{\left( {\mathcal{E} - {\Omega _ + }\mathcal{J} + \frac{{e q {r_+ }}}{{r_+ ^2 + {a^2}}}} \right)\pi i}}{{2\kappa }} \, .
\label{formula46}
\ee
The tunneling probabilities of the outgoing and ingoing particle are \cite{Srinivasan:1998ty,Shankaranarayanan:2000qv}
\be
P_{emission} \sim e^{-2 \textrm{Im} I} = e^{-2 \textrm{Im} W_{+} }\, , \label{formula47}
\ee
\be
P_{absorption} \sim e^{-2 \textrm{Im} I} = e^{-2 \textrm{Im} W_{-}}\, . \label{formula48}
\ee
Since $Im W_{+}=- Im W_{-}$, the probability of the particle tunneling from inside to outside the horizon takes the form
\be
\Gamma \sim P_{emission}/P_{absorption} = {e^{ - 4{\textrm{Im}}{W_+ }}}\, . \label{formula49}
\ee
Eventually, we get the tunneling probability of the charged complex scalar particle
\be
\Gamma \sim {e^{ - 4{\textrm{Im}}{W_+ }}}
= \exp \left[ { - \frac{{2\pi }}{\kappa }\left( \mathcal{E} - {\Omega _ + }\mathcal{J} + \frac{e q {r_+}}{r_+^2 + a^2} \right)} \right].
\label{formula50}
\ee
Obviously, neglecting the varied background spacetime, the tunneling rate derived here differs from that in (\ref{formula36}).

%%%%%%%%%%%%%%%%%%%%%%%%%%%%%%%%%%%%%%%%%%%%%%%%%%%%%%%%%%%%%%%%%%%%%%%%%%%%%%%%%%%%%
\section{Tunneling radiation of charged Dirac particles}
%%%%%%%%%%%%%%%%%%%%%%%%%%%%%%%%%%%%%%%%%%%%%%%%%%%%%%%%%%%%%%%%%%%%%%%%%%%%%%%%%%%%%
\label{TPOCDP}

In this section, we investigate the tunneling radiation of charged spin-$1/2$ Dirac particles. Our treatment differs from that in
\cite{Kerner:2006vu,CWY1,CJWY,CWY2} and is more convenient than the approaches in \cite{Rehman:2010zs,Jiang:2008gq} etc., because
we just need to deal with the radial Dirac equations for calculating the tunneling probability.

Setting units $G = \hbar  = c = 1$, the Dirac equation in curved background spacetime can be written as
\be
\big({\mathbb{H}_D} + {\mu_e}\big)\Psi = \big[{\gamma^A}e_A^{\mu }({\p_\mu } + {\Gamma _\mu } + iq_1\tilde{{\mathcal{A}}_\mu }) + {\mu_e}\big]\Psi  = 0,
\label{formula51}
\ee
where $\mu_e$ and $q_1$ refer to the mass and electric charge of particles, the f\"{u}nfbein $e_a^{~\mu}$ and its inverse $e_{~\mu}^a$ are
defined by the spacetime metric $g_{\mu\nu} = \eta_{ab}e_{~\mu}^ae_{~\nu}^B$ with $\eta_{ab} = diag (-1, 1, 1, 1)$ being the flat (Lorentz)
metric tensor, Latin letters $a, b$ denote local orthonormal (Lorentz) frame indices  $\{0, 1, 2, 3\}$, while Greek letters $\mu, \nu$
run over four-dimensional spacetime coordinate indices $\{t, r, \theta, \phi\}$. Then we proceed to choose gamma matrices $\gamma^a$
obeying the anticommutation relations (Clifford algebra)
\be
\big\{\gamma^a, \gamma^b\big\} \equiv \gamma^a\gamma^b +\gamma^b\gamma^a = 2\eta^{ab} \, ,
\label{formula52}
\ee
and take an explicit representation of the Clifford algebra as following,
\bea
&& \gamma^0 = i\Big(\ba{cc}
 0 & I \\
 I & 0 \ea \Big) \, , \qquad
\gamma^1 = i\Big(\ba{cc}
 0         & \sigma^3 \\
 -\sigma^3 & 0 \ea \Big) \, , \qquad \nn \\
&& \gamma^2 = i\Big(\ba{cc}
 0         & \sigma^1 \\
 -\sigma^1 & 0 \ea \Big) \, , \qquad
  \gamma^3 = i\Big(\ba{cc}
 0         & \sigma^2 \\
 -\sigma^2 & 0 \ea \Big) \, , \qquad
\label{formula53}
\eea
where $\sigma^i$ are the Pauli matrices, and $I$ is a  $2 \times 2$ identity matrix, respectively.

Similarly to the analysis in \cite{Wu:2008df}, after calculating the spin-connection 1-form $\omega_{ab}$ in the orthonormal frame and
constructing the spinor connection 1-form $\Gamma$, in terms of the local differential operator $\p_b = e_a^{~\mu}\p_{\mu}$, the Dirac
equation (\ref{formula51}) can be rewritten in the local Lorentz frame as
\be
\Big[\gamma^a\left(\p_a +\Gamma_a + iq_1 \tilde{\mathcal{A}}_a\right) +\mu_e\Big]\Psi = 0,
\label{formula54}
\ee
where $\Gamma_a = e_a^{~\mu}\Gamma_{\mu} = (1/4)\gamma^b\gamma^cf_{bca}$ is the component of the spinor connection in the local Lorentz
frame.

With regard to the new form of the four-dimensional charged rotating AdS black holes solution (\ref{formula01}), we employ the following
local Lorentz basis of 1-forms (tetrad) according to the definition that $e^a = e^a_{~\mu}dx^{\mu}$ orthonormal with respect to $\eta_{ab}$,
\bea
{e^0} &=& \frac{1}{\Xi }\sqrt {\frac{{{\Delta _r}}}{\Sigma }} ({\Delta _\theta }d\tilde{t} - a{\sin ^2}\theta d\tilde{\phi} )\, , \quad
{e^1} = \sqrt {\frac{\Sigma }{{{\Delta _r}}}} dr \, \nn \\
{e^2} &=& \sqrt {\frac{\Sigma }{{{\Delta _\theta }}}} d\theta \, , \quad
{e^3} = \frac{{\sin \theta }}{\Xi }\sqrt {\frac{{{\Delta _\theta }}}{\Sigma }} \left[({r^2} + {a^2})d\tilde{\phi}  - a(1 + {g^2}{r^2})d\tilde{t}\right]. \quad~
\label{formula55}
\eea
With the above local Lorentz basis of 1-forms in hand, we proceed to present the orthonormal basis 1-vectors $\p_a = e_a^{~\mu}\p_{\mu}$ dual
to the tetrad $e^{a}$,
\bea
\p_0 &=& \frac{1}{{\sqrt {{\Delta _r}\Sigma } }}\bigg[({r^2} + {a^2}){\p_{\tilde{t}}} + a(1 + {g^2}{r^2}){\p_{\tilde{\phi}} }\bigg]\, , \quad
\p_1  = \sqrt {\frac{{{\Delta _r}}}{\Sigma }} {\p_r}\, , \nn \\
\p_2 &=& \sqrt {\frac{\Delta _\theta}{\Sigma }} {\p_\theta}\, , \qquad
\p_3 = \frac{1}{\sin\theta \sqrt {{\Delta _\theta}\Sigma } }\bigg(\Delta _\theta {\p _{\tilde{\phi}}} + a{\sin^2}\theta {\p_{\tilde{t}}}\bigg).
\label{formula56}
\eea
After working out ${\gamma^a}{\p_a}$, $\gamma^a\Gamma_a$ and the coupling term ${\gamma^a}{\tilde{\mathcal{A}}_a}$, it's easy to obtain the
Dirac equation. To facilitate calculating the tunneling probability, we could assume following ansatz for variables separation
\be
\Psi = e^{i(m_2\phi-\omega_2 t)}\left(\ba{cl}
&\hspace*{-5pt} R_2(r)S_1(\theta) \\
&\hspace*{-5pt} R_1(r)S_2(\theta) \\
&\hspace*{-5pt} R_1(r)S_1(\theta) \\
&\hspace*{-5pt} R_2(r)S_2(\theta)
\ea\right) \, .
\label{formula57}
\ee
As a result, the Dirac equation can be written in decoupled form. The radial part is
\be
\hspace{-1.5cm} \left[ -\sqrt{\Delta_r} {D_r} +i\frac{\omega_2\left(r^2 +a^2\right)
-a m_2\left(1 +g^2r^2 \right) +q q_1 r}{\sqrt{\Delta_r}} \right]{R_1}
 = \left( \lambda  +\frac{i{\mu _e}r}{\sqrt{2}} \right){R_2},
\label{formula58}
\ee
\be
\hspace{-1.5cm} \left[ -\sqrt{\Delta_r} {D_r} -i\frac{\omega\left(r^2 +a^2\right)
-a m_2\left(1 +g^2r^2 \right) +q q_1 r}{\sqrt{\Delta_r}} \right]{R_2}
 = \left( \lambda  -\frac{i{\mu _e}r}{\sqrt{2}} \right){R_1},
\label{formula59}
\ee
and the angular part is
\bea
&&\hspace*{-1cm}
\left[{\Delta_\theta}{L_\theta} -\frac{m_2\Delta_\theta +a\omega_2{\sin^2\theta}}{\sin\theta \sqrt{\Delta_\theta}}\right]{S_1}
= \left(\lambda  -\frac{a{\mu_e}{\cos\theta}}{\sqrt{2}}\right){S_2},  \\
\label{formula60}
&&\hspace*{-1cm}
\left[{\Delta_\theta}{L_\theta} +\frac{m_2\Delta_\theta +a\omega_2{\sin^2\theta}}{\sin\theta \sqrt{\Delta_\theta}}\right]{S_2}
= \left(\lambda  +\frac{a{\mu_e}{\cos\theta}}{\sqrt{2}}\right){S_1},
\label{formula61}
\eea
where we have introduced two operators
\be
D_r = \p_r +\frac{\Delta_r^\prime}{4\Delta_r}\, ,\qquad
L_\theta = \p_\theta +\frac{\Delta_\theta^\prime}{4\Delta_\theta} + \frac{1}{2}\cot\theta\, .
\label{formula62}
\ee

Next, we focus our attention on the tunneling probability of the Dirac particles across the charged rotating AdS black holes in conformal
gravity. It is worth pointing out that there is only need to deal with the radial equation while ignoring the angular part. As for the
spin-$1/2$ Dirac particles, when measuring spin along $z$ direction, there would be two cases including spin up and spin down. Here we
only discuss the case of spin up in detail. The process of the spin down case is parallel. After recovering $\hbar$ and taking place
$m_2,\omega_2$ with $\frac{\mathcal{J}}{\hbar},\frac{\mathcal{E}}{\hbar}$ in Eqs.(\ref{formula58}) and (\ref{formula59}), we substitute
the variables $R_1,R_2$ for
\be
R_1 = Ae^{\frac{i}{\hbar}{H\left( r \right)}}\, , \qquad R_2 = Be^{\frac{i}{\hbar}{H\left( r \right)}}\, . \quad
\label{formula63}
\ee
Substituting equation (\ref{formula63}) into (\ref{formula58}), (\ref{formula59}) and neglecting high order term ${\Delta_r^\prime}/(4\Delta_r)$, we get
\be
\hspace{-2.0cm} H^\prime\left( r \right) = \pm \frac{1}{\Delta_r}\Big\{\left[ {\mathcal{E}\left( {{r^2} + {a^2}} \right) -
a\mathcal{J}\left( {1{\rm{ + }}{{\rm{g}}^2}{r^2}} \right) + q q_1 r} \right]^2
 -{\Delta_r}{\hbar^2}\big( \lambda  -\frac{i{\mu_e}r}{\sqrt{2}} \big) \Big\}^{1/2}\, , \quad
\label{formula64}
\ee
where the sign $``\pm"$ corresponds to the cases of ingoing and outgoing particles. Taking into account the asymptotic behaviors of particles
as well as the angular velocity $\Omega_+$ near the event horizon in equation (\ref{formula44}), we integrate $H^\prime\left( r \right)$ yields
\be
H\left(r \right) = \frac{\pi i}{2\kappa}\left(\mathcal{E} -{\Omega_+}\mathcal{J} + \frac{q q_1 {r_+}}{r_+^2 + a^2} \right).
\label{formula65}
\ee
According to the analysis of Srinivasan et al in Refs.\cite{Srinivasan:1998ty,Shankaranarayanan:2000qv}, the tunneling probability of charged spin-$1/2$
Dirac particles is give by
\be
\Gamma \sim e^{ - 4\textrm{Im}{H(r)}}
= \exp \left[ { - \frac{{2\pi }}{\kappa }\left( {\mathcal{E} - {\Omega _ + }\mathcal{J} + \frac{{q q_1 {r_ + }}}{{r_+^2 + {a^2}}}} \right)} \right].
\label{formula66}
\ee
From Eqs. (\ref{formula50}) and (\ref{formula66}), we find that both the complex scalar particles and spin-$1/2$ Dirac particles
with the same charge tunneling across the charged rotating AdS black holes in conformal gravity share the same probability.

%%%%%%%%%%%%%%%%%%%%%%%%%%%%%%%%%%%%%%%%%%%%%%%%%%%%%%%%%%%%%
\section{conclusions and discussion}
%%%%%%%%%%%%%%%%%%%%%%%%%%%%%%%%%%%%%%%%%%%%%%%%%%%%%%%%%%%%%
\label{CADA}

Researches of Hawking radiation in different gravity theories is of important significance.  In this paper, we have extended to investigate the tunneling
radiation of the charged rotating AdS black hole in conformal gravity for the charged massive particles, the complex scalar and spin-$1/2$ Dirac particles.
Previous works which studied Hawking radiation in the rotating spacetime used to depend upon the dragging coordinate transformation. In our work, the
coordinate transformation has been avoided. The research on the rotating black holes' tunneling radiation is not limited in a dragging coordinate system any
more. What's more, with remedying shortcomings in deriving the geodesic equation of particles, we have derived the massive and massless particles' equation
of motion in a unified and self-consistent way.

Finally, for the case of the charged massive particle, Parikh and Wilczek's result was recovered. It is noteworthy that, the entropy of the charged rotating
AdS black holes in conformal gravity is not one-quarter of the horizon area, which is different from the case in Einstein gravity theory, nevertheless the
tunneling probability of the charged massive particle is still related to the change of Bekenstein-Hawking entropy. But, neglecting the varied background
spacetime, both of the tunneling probabilities of the complex scalar particle and the spin-$1/2$ Dirac particle is not directly related to the entropy variation.

As an intriguing higher-derivative theory, conformal gravity shares more properties which need to be further studied. That would be left us for future discussions.

%%%%%%%%%%%%%%%%%%%%%%%%%%%%%%%%%%%%%%%%%%%%%%%%
\section{Acknowledgments}
%%%%%%%%%%%%%%%%%%%%%%%%%%%%%%%%%%%%%%%%%%%%%%%%

I appreciate Professor Shuang-Qing Wu for helpful discussion. I would also like to give great thanks to the referees for their helpful
advices. This work is supported by the National Natural Science Foundation of China with Grant Nos. 11275157 and 11205125.

\end{document}